\documentclass[12pt]{article}
\usepackage{graphicx}
\usepackage{amssymb}
\usepackage{wrapfig}
\usepackage{amsmath}

\makeatletter
\@addtoreset{equation}{section}
\makeatother

\newcommand{\VEV}[1]{\left\langle #1 \right\rangle}

\newcommand{\diag}{{\rm diag}}

\newcommand{\bequ}{\begin{equation}}
\newcommand{\eequ}{\end{equation}}
\newcommand{\beqn}{\begin{eqnarray}}
\newcommand{\eeqn}{\end{eqnarray}}
\newcommand{\bctr}{\begin{center}}
\newcommand{\ectr}{\end{center}}
\newcommand{\bit}{\begin{itemize}}
\newcommand{\eit}{\end{itemize}}

\begin{document}
\begin{titlepage}

\begin{flushright}
DPNU-06-07\\
\today
\end{flushright}

\vspace{4ex}

\begin{center}
{\large \bf
A Solution for Little Hierarchy Problem and $b\rightarrow s \gamma$
}

\vspace{6ex}

\renewcommand{\thefootnote}{\alph{footnote}}
S.-G. Kim\footnote{e-mail: sunggi@eken.phys.nagoya-u.ac.jp}, 
N. Maekawa\footnote{e-mail: maekawa@eken.phys.nagoya-u.ac.jp},
A. Matsuzaki\footnote{e-mail: akihiro@eken.phys.nagoya-u.ac.jp}, 
K. Sakurai\footnote{e-mail: sakurai@eken.phys.nagoya-u.ac.jp}, \\
A. I. Sanda\footnote{Now at Department of Engineering, Kanagawa University. e-mail: sanda@kanagawa-u.ac.jp},
 and 
T. Yoshikawa\footnote{e-mail: tadashi@eken.phys.nagoya-u.ac.jp}

\vspace{4ex}
{\it Department of Physics, Nagoya University, Nagoya 464-8602, Japan}\\

\end{center}

\renewcommand{\thefootnote}{\arabic{footnote}}
\setcounter{footnote}{0}
\vspace{6ex}

\begin{abstract}
We show that all the parameters which destabilize the weak scale can be taken around the weak scale in the MSSM
without conflicting with the SM Higgs mass bound set by LEP experiment. 
The essential point is that if the lightest CP-even Higgs $h$ in the MSSM has only a small coupling to 
$Z$ boson, $g_{ZZh}$,
LEP cannot generate the Higgs sufficiently. In the scenario, the SM Higgs mass bound 
constrains the mass of
the heaviest CP-even Higgs $H$ which has the SM like $g_{ZZH}$ coupling. However, it is easier to make the heaviest
Higgs heavy by the effect of off-diagonal elements of the mass matrix of the CP-even Higgs because the larger eigenvalue
of $ 2\times 2$ matrix becomes larger by introducing off-diagonal elements. Thus, the smaller stop masses can be
 consistent with the LEP constraints. Moreover, the two excesses observed at LEP Higgs search can naturally be
 explained as the signals of the MSSM Higgs $h$ and $H$ in this scenario. 
   
One of the most interesting results in the scenario is that all the Higgs in the MSSM have the weak scale masses.
For example, the charged Higgs mass should be around 130 GeV. This looks inconsistent with the lower bound obtained
by the $b\rightarrow s \gamma$ process as $m_{H^\pm}>350$ GeV. 
However, we show that the amplitude induced by the charged Higgs can naturally be compensated by that of the chargino if we take the mass parameters by which the little hierarchy problem can be solved. 
The point is that the both amplitudes have the same order of magnitudes when all the fields in the both loops have the same order of masses. 
\end{abstract}

\end{titlepage}


\section{Introduction}

Supersymmetry (SUSY) is one of the most promising solutions for the weak scale instability of the standard model (SM)
. In the minimal supersymmetric standard model (MSSM), 
there are several additional attractive points. Three gauge 
couplings meet at a scale, which
strongly implies the SUSY grand unified theory (GUT). And the lightest SUSY particle (LSP) becomes
stable, which can be a dark matter candidate.

However, this simple model seems to be unsatisfactory.
The problem is related to one of the characteristic features of the MSSM; the mass of the lightest CP-even Higgs boson ($h$) is always smaller than the Z boson mass at tree level\footnote
{Here, we take usual notations for CP-even Higgs bosons $ h $ and $ H $, i.e., the mass of $h$ is always smaller than that of $H$. CP violation in the Higgs sector which causes the mixing of $ h $, $ H $, and CP-odd Higgs boson $ A $ is ignored in the following discussions.}.
However, LEP2 experiment gives us severe bound for the SM Higgs as $m_{\varphi_{\rm SM}}>114.4$ 
GeV (95" C.L.)\cite{Barate:2003sz} .
This lower bound for the Higgs mass is not inconsistent with the MSSM prediction if loop corrections to the Higgs 
potential are taken into account\cite{MSSMHiggs}.
The largest contribution is induced by stop loop correction as
\begin{eqnarray}
m_h^2&\leq& m_Z^2+\Delta_{22}, \\
\Delta_{22}&\sim&\frac{3Y_t^4\VEV{H_u}^2}{4\pi^2}\log\frac{m_{\tilde t}^2}{m_t^2},
\end{eqnarray}
where $Y_t$, $m_{\tilde t}$, and $m_t$ are the top Yukawa coupling, the stop mass, and the top mass,
 respectively.
Because it is a logarithmic function of the stop mass, 
if the lower mass bound of the SM Higgs, $m_{\varphi _{\rm SM}}>114.4$ GeV 
is naively applied to the mass of $h$, then the stop mass has to be larger 
than 500 GeV.
On the other hand, the stop also contributes to the mass parameter of up-type Higgs field $H_u$ as
\begin{eqnarray}
m_{H_u}^2&=&m_{H_u0}^2+\Delta m_{H_u}^2,\\
\Delta m_{H_u}^2&\sim& -\frac{3Y_t^2}{4\pi^2}m_{\tilde t}^2\log\frac{\Lambda}{m_{\tilde t}}.
\end{eqnarray}
Such a large stop mass leads to a tuning between the tree Higgs mass parameter 
$m_{H_u0}^2$ and the correction $\Delta m_{H_u}^2$, because in order to obtain the correct weak boson masses, the renormalized Higgs mass parameter 
$m_{H_u} $ must be around the weak scale, $O(m_W)$.
For example, if the cutoff scale $\Lambda$ is taken as the Planck scale, 
previous requirement for stop mass results in  $|\Delta m_{H_u}^2|\geq (850{\rm GeV})^2$ and less than a percent fine-tuning is required in this naive analysis.
This difficulty is called as ``the little hierarchy problem" and in the literatures, various solutions have been
examined\cite{little1}-\cite{little13}.

However, the LEP bound cannot be applied directly if the Higgs sector is extended, e.g., 
to two Higgs doublet models (2HDM) as in the case of the MSSM.
The main production mode of the Higgs at LEP experiment is $e^+e^-\rightarrow Z^*\rightarrow Z h$.
Therefore, if the coupling of the lightest CP-even Higgs $h$ to Z boson, $g_{ZZh}$, is sufficiently 
smaller than that of the SM Higgs to Z boson, $g_{ZZ\varphi _{\rm SM}}$, LEP cannot produce $h$ sufficiently and there is no need to treat the LEP constraint as a lower mass bound of $h$.
In fact, the report from the LEP Working Group for Higgs searches \cite{Barate:2003sz} roughly gives the 
bound for the $g_{ZZh}$ coupling as $g_{ZZh}<g_{ZZ\varphi_{\rm SM}}/2$ for the Higgs with $m_h>90$ GeV.
Moreover, it was reported that there were 2.3 and 1.7 $\sigma$ excesses from the back ground estimations of 
Higgs search experiment at the corresponding Higgs boson mass with nearly 98 and 115 GeV, respectively.
Notably, the former excess is too small to identify it as productions of the SM Higgs with the mass of $98$ GeV, but it can be explained by the MSSM Higgs $h$ with small $g_{ZZh}$ coupling.


In this article, we regard this small $g_{ZZh}$ scenario as a way to bypass the little hierarchy problem.
%
It has already been pointed out that the two excesses observed at LEP experiment can be explained by the MSSM Higgs in the literatures \cite{Kane:2004tk,Drees:2005jg}.
However, it is not obvious whether the little hierarchy problem can be solved simultaneously, or, natural scenario is possible in such a situation, because in their numerical calculation
they allow large SUSY breaking parameters (stop masses, $A$ parameters, gaugino masses) and large SUSY Higgs mass parameter $\mu$, which should preferably be around the weak scale in order to avoid the tuning problem.
Therefore, in this paper, we examine a scenario of light $h$ with small $g_{ZZh}$ coupling taking the SUSY 
breaking parameters and $\mu$ as low as its experimentally allowed values and show that a large parameter region can be consistent with the LEP experiment, even if we assume stop mass is smaller than $500$ GeV.
After this introduction, in section 2, we recall how the Higgs with small $g_{ZZh}$ coupling can
be obtained and show the essential point for this scenario. And in section 3, we present the 
numerical calculation and show that such a natural scenario is possible. 
Moreover, in section 4, even though the scenario of small $g_{ZZh}$ coupling predicts all the MSSM Higgs 
masses to be $O(100\,{\rm GeV})$, we will show that such a light charged Higgs  which looks inconsistent
 with the $b\rightarrow s\gamma$ constraint (e.g., $m_{H^\pm} \geq 350$ GeV in the case of type II 2HDM \cite{Gambino:2001ew} ) at first glance is admissible because of 
a cancellation between the charged Higgs contribution and the chargino's.
The requirement for solving the little hierarchy problem play an essential role for the cancellation.



\section{The lightest Higgs with small $g_{ZZh}$ coupling}
We demonstrate how the lightest CP-even Higgs $h$ with small $g_{ZZh}$ coupling constant consistent with the 
 LEP constraint can be realized and accompany light stop.
We also clarify the essential points.

First of all, $g_{ZZHiggs}$ coupling originates from the $ZZH^\dagger H$ interaction, 
substituting a vacuum expectation value (VEV) for one of the Higgs fields.
Therefore, this coupling constant is proportional to the VEV of the corresponding Higgs field.
In the models with two Higgs doublets as in the case of the MSSM, we can take generally linear combinations of two Higgs doublets as, $h_{VV}$, which has a vanishing VEV, and the other combination, $h_{SM}$, has a VEV whose value equals to that of the SM Higgs field.
They are written as

\begin{equation}
\begin{pmatrix}
h_{VV} \\ h_{SM}
\end{pmatrix} 
=
\begin{pmatrix} 
 \sin\beta& -\cos\beta \\ \cos\beta&\sin\beta
\end{pmatrix}
\begin{pmatrix} 
H_d \\ H_u
\end{pmatrix},
\end{equation}
where $H_d$ is the down-type Higgs field and $\tan\beta\equiv \VEV{H_u}/\VEV{H_d}$.
We take $\cos\beta$ and $\sin\beta$ as positive value.
It is obvious that $h_{VV}$ has vanishing $g_{ZZh_{VV}}$ coupling because its VEV is vanishing. 
Therefore, if the main mode of the lightest CP-even Higgs boson $h$ is $h_{VV}$, such $h$ becomes invisible at LEP experiments.

Next we show the main mode of the lightest CP-even Higgs boson can really be $h_{VV}$ in the MSSM.
For simplicity, here, we use tree level Higgs potential of the MSSM\footnote{
This is sufficient for the following qualitative arguments. 
Later on, we will use one-loop effective potential, when we discuss the scenario quantitatively.};
\begin{multline}
V_{\rm tree}=m_1^2|H_d|^2+m_2^2|H_u|^2+(m_3^2H_uH_d+h.c.) \\
+\frac{g^2}{8}\left(H_u^\dagger\tau^a H_u+H_d^\dagger\tau^a H_d\right)^2
+\frac{g'^2}{8}\left(H_u^\dagger H_u-H_d^\dagger H_d\right)^2,
\end{multline}
where $\tau^a$ $(a=1,2,3)$ are the Pauli matrices.
Four point couplings of the MSSM Higgs potential is written by $SU(2)_L$ and $U(1)_Y$ gauge couplings,
$g$ and $g'$.
Therefore, two point couplings, $m_i^2$ (i=1,2,3), are the parameters of this potential.
One freedom of three parameters is fixed by the weak boson masses,
only two parameters determine the potential completely.
In the following, we use CP-odd Higgs boson mass, $m_A$, and $\tan\beta$ as the two parameters.
It is straightforward to calculate the CP-even Higgs mass matrix as
\begin{equation}
M_{h^0}=\bordermatrix{& H_d & H_u \cr
                   H_d & m_A^2\sin^2\beta+m_Z^2\cos^2\beta & -(m_A^2+m_Z^2)\sin\beta\cos\beta \cr
                   H_u &  -(m_A^2+m_Z^2)\sin\beta\cos\beta & m_Z^2\sin^2\beta+m_A^2\cos^2\beta}.
\end{equation}
Here $m_A^2=m_1^2+m_2^2$ at this tree level approximation.
Using this matrix, we show the condition which makes lightest CP-even Higgs boson as $h_{VV}$. 
For simplicity, we take $\tan\beta \gg 1$, namely, $\cos\beta\sim 0$.
Then, the mass matrix becomes diagonal form, 
$M_{h^0}\sim\diag(m_A^2, m_Z^2)$ 
and from the previous discussion, these diagonal entries correspond to the mass of $H_d \sim h_{VV}$ and $H_u \sim h_{\rm SM}$, since $\tan\beta \gg 1$ means $H_u$ gets almost the same VEV as the SM Higgs field and $\VEV{H_d} \approx 0$.
Therefore if we take $m_A<m_Z$, we can obtain the lightest CP-even Higgs $h$ with small $g_{ZZh}$ 
coupling.
We call this situation as ``Inverse case" ($m_A < m_Z$ in tree level ) for the later convenience and also name ``Normal case" for the situation where up-type Higgs mass is lighter than down-type Higgs mass 
($m_A > m_Z$ in tree level). Note that in the Inverse case, all the mass scales $m_A$ and $m_Z$ are around
the weak scale, which means that all the Higgs masses must be around the weak scale. 

In practice, however, when off-diagonal components are neglected, i.e., $\cos \beta \sim 0$, there is no difference between the ``Normal case" and the ``Inverse case" respect to the mass of $H_u$, 
since it is independent of $m_A$.
Even if we consider the one loop corrected mass square only for the up-type Higgs field as, 
\begin{equation}
M_{h^0}\sim\bordermatrix{& H_d & H_u \cr
                   H_d & m_A^2 & -(m_A^2+m_Z^2)\sin\beta\cos\beta \cr
                   H_u &  -(m_A^2+m_Z^2)\sin\beta\cos\beta & m_Z^2+\Delta_{22}}
=
\begin{pmatrix}
a&c \cr
 c&b
\end{pmatrix},                
\label{eq:corrected_mass_matrix}
\end{equation}
the mass of $H_u$ in the ``Inverse case" is the same as in the ``Normal case".
But once we take into account the off-diagonal entries, 
we can show that in the ``Inverse case", smaller stop mass is sufficient to meet the LEP bound comparing to the ``Normal case".
This is reflections of general features of diagonalization of a matrix.
That is, if there are off-diagonal entries, the larger (smaller) eigenvalue becomes always larger (smaller) than 
the larger (smaller) diagonal element.
(Actually, the eigenvalues of $2\times 2$ matrix become $x_\pm=(a+b\pm\sqrt{(a-b)^2+4c^2})/2$.)
Therefore, in the ``Inverse case", the larger eigenvalue ($m_H^2$) which has to satisfy the LEP constraint 
increases by introducing off-diagonal element and the constraint becomes milder.
On the contrary, 
in the ``Normal case", the smaller eigenvalue ($m_h^2$) which has to satisfy the LEP constraint decreases
by introducing off-diagonal element and the constraint becomes stronger.
Thus, in the ``Inverse case", the smaller stop mass can be sufficient for the LEP constraint than that 
in the ``Normal case". 
This is the essence of the scenario which can open the way to ease the fine-tuning.
It is obvious that the larger off-diagonal component requires larger stop mass in the 
``Normal case", while smaller stop mass becomes sufficient in the ``Inverse case".
Since larger $\tan\beta$ leads to smaller off-diagonal component, larger $\tan\beta$ is preferable for the ``Normal case", while smaller $\tan\beta$ is preferable for the ``Inverse case".


\section{Numerical analyses}
We explore the scenario which realizes the Inverse case discussed in the previous section numerically.

First of all, we explain how to determine the SUSY-breaking parameters and $\mu$ parameter 
used in the following analyses.
Since large values for these parameters entail tuning problem, 
we take each parameters as low as they don't conflict its own experimental constraint.

\begin{itemize}

\item Gaugino masses : $M_a$ {\small $(a=1,2,3)$} and SUSY invariant Higgs mass : $\mu$

We assume GUT relations for gaugino masses.
We take $M_1=60$ GeV, $M_2=120 $ GeV, $M_3=400$ GeV, and $\mu=250$ GeV at the weak scale, which
satisfy the present experimental constraints ($m_{\chi ^0}>46$ GeV and $m_{\chi ^{\pm }}>94$ GeV \cite{PDG05})
for a large region of $\tan\beta$.
These mass parameters correspond to the gaugino mass $M_{\frac{1}{2}}\sim 145$ GeV and 
$\mu\sim 250$ GeV at the GUT scale.

\item Stop masses : $m_{\tilde t_{L}}$ and $m_{\tilde t_{R}}$

We assume universal soft masses ($m_0$) for squarks and sleptons at the GUT scale.
The stop masses at the weak scale can be determined by the gaugino masses and the scalar
masses. 
Because current lower slepton mass bound ($m_{\tilde \tau }>81.9$ GeV \cite{PDG05}) determines the scale of $m_0$  as $m_0\gtrsim 80$ GeV, we set $m_0\sim 100$ GeV and then we take the left and right-handed stop masses 
at the weak scale as $m_{\tilde t_{L}}=350$ GeV and $m_{\tilde t_{R}}=300$ GeV, respectively.



\item Scalar three point coupling : $A_X$ {\small $(X=U,D,E)$}

We assume each $A_X$ is proportional to the corresponding Yukawa coupling with uniform factor at the GUT scale 
($A_X=A Y_X$). We set $A_t=$ 300 GeV, 325 GeV, and 350 GeV at the weak scale as typical values.
These values correspond to the gaugino mass $M_{\frac{1}{2}}\sim 145$ GeV and the universal $A$ parameter 
$A\sim 0$, $A\sim 125$ GeV, and $A\sim 250$ GeV at the 
GUT scale, respectively. Under this universal $A$ parameter assumption, $A$ larger than 250 GeV can induce 
the charge breaking at the weak scale. 

\item Higgs mass parameters : $m_1, \, m_2, \, m_3$

We assume no constraints for these three parameters at GUT scale and treat two of them ($m_A$ and $\tan\beta $) as free at the weak scale. 
\end{itemize}
Here, we took the several values for $A_t$, but not for $m_{\tilde t}$. This is because the results are more sensitive
to the parameter $A_t$ than the stop masses if we consider the naturalness seriously.

%

\begin{figure}[t]
\begin{center}
\includegraphics[width=.7\textwidth]{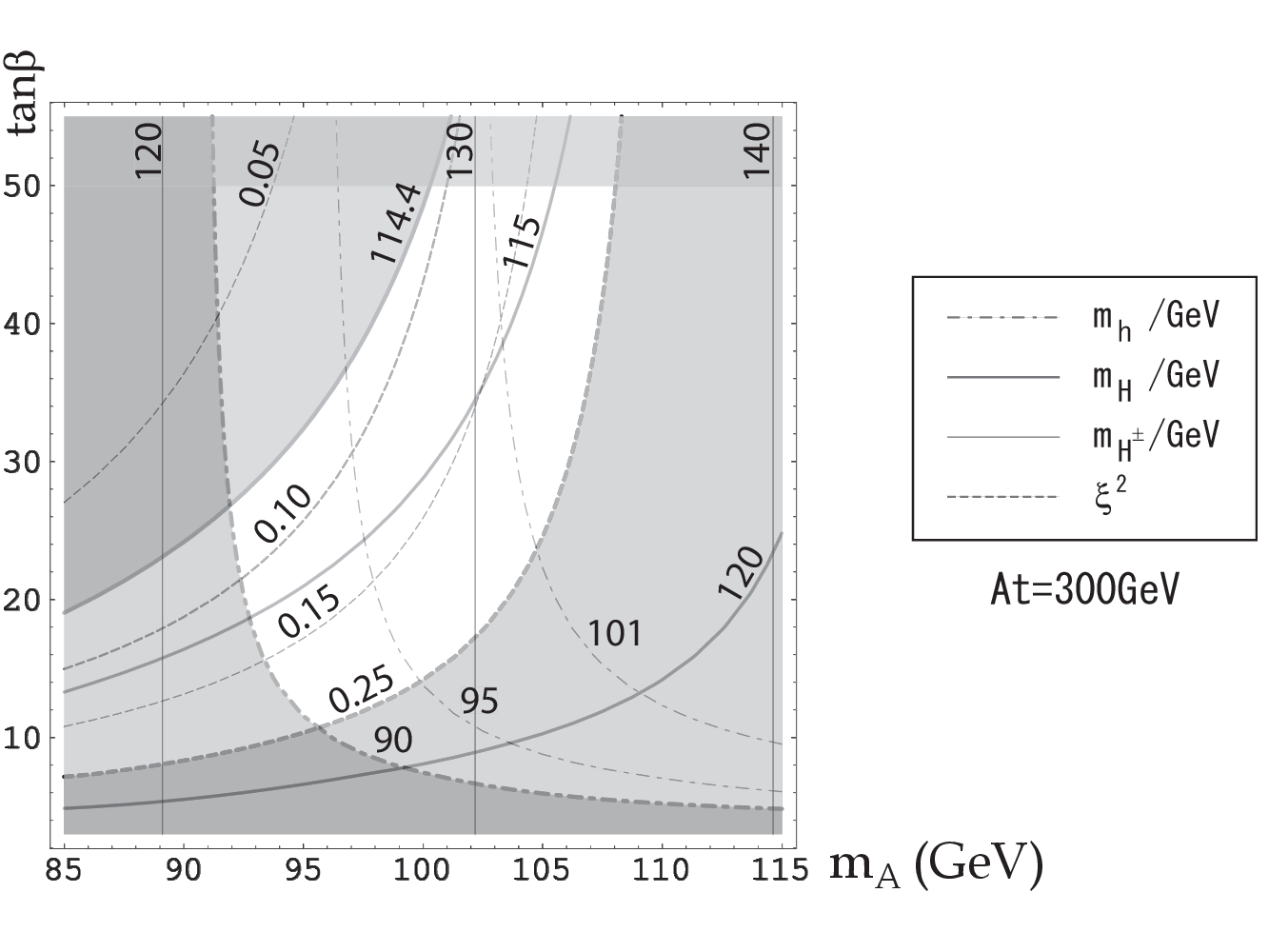}
\caption{ 
{\footnotesize
Doted-dashed lines represent contour lines of $m_h=90$, 95, and 101 GeV. Dashed lines represent contour lines of $\xi^2 \equiv (g_{ZZh}/g_{ZZ\varphi_{\rm SM}})^2= 0.05$, 0.10, 0.15, and 0.25. Thick solid lines represent contour lines of
$m_H=$114.4, 115, and 120 GeV and thin solid lines represent contour lines of $m_{H^\pm}=120$, 130, and 140 GeV. We set
$A_t$=300 GeV.
White area is the allowed region.
}}
\label{inverse_a}
\end{center}
\end{figure}

\begin{figure}[t]
\begin{center}
\includegraphics[width=.7\textwidth]{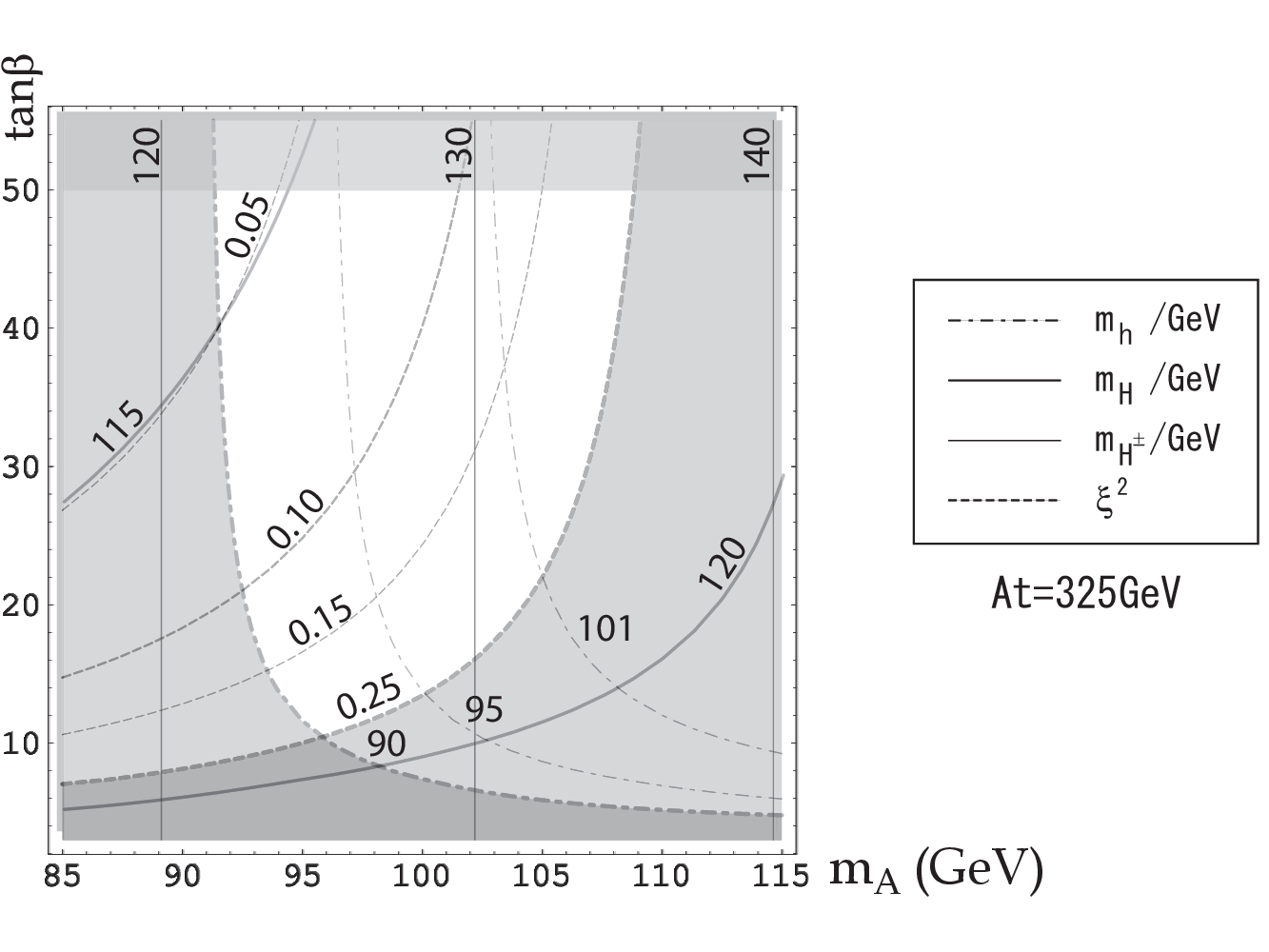}
\caption{ 
{\footnotesize
We set $A_t=325$ GeV.
Each line represents contour line as in the case of Fig.\ref{inverse_a}.
White area is the allowed region.
}}
\label{inverse_b}
\end{center}
\end{figure}

\begin{figure}[t]
\begin{center}
\includegraphics[width=.7\textwidth]{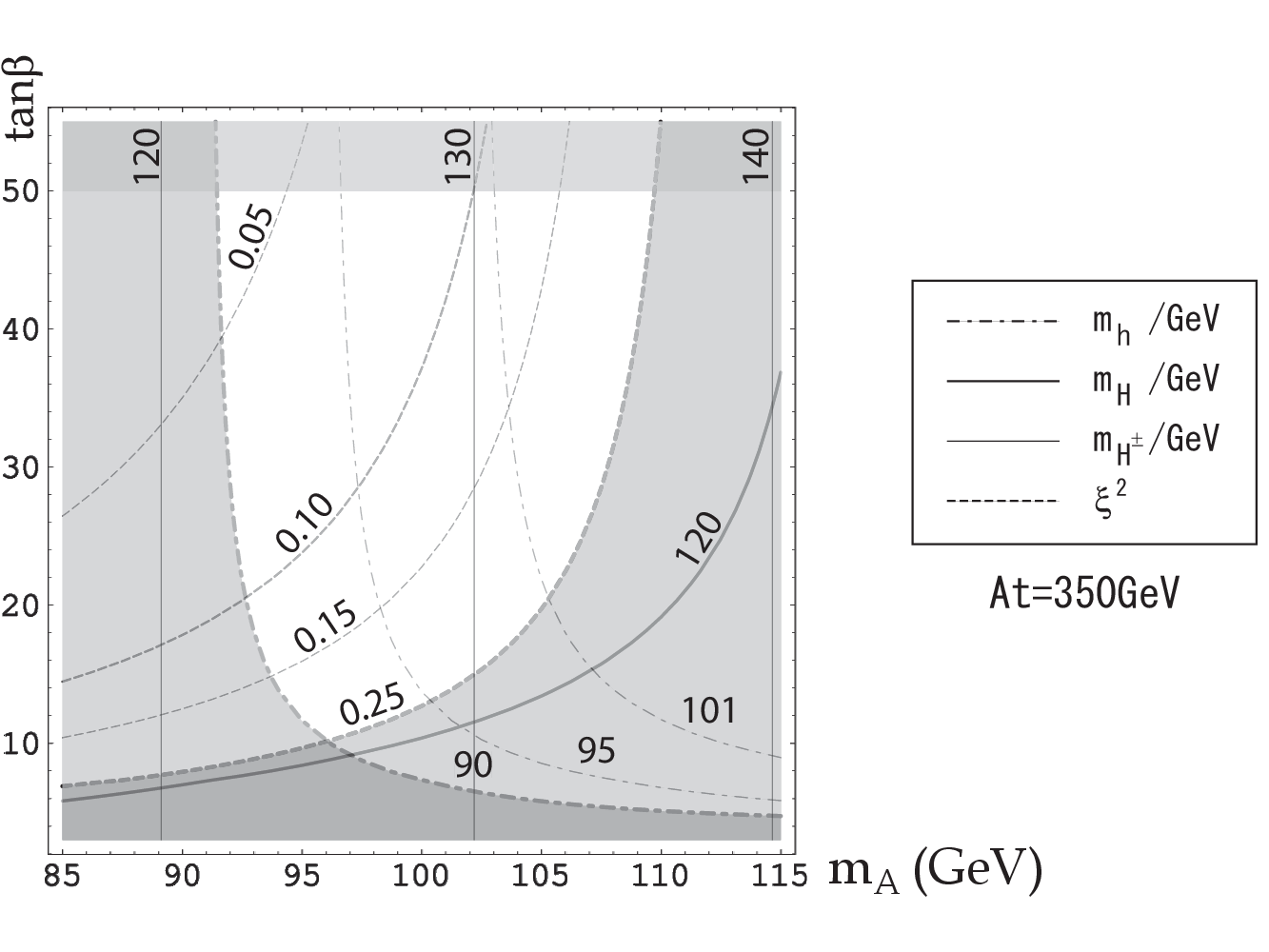}
\caption{ 
{\footnotesize
We set $A_t=350$ GeV.
Each line represents contour line as in the case of the previous figures.
White area is the allowed region.
}}
\label{inverse_c}
\end{center}
\end{figure}

Let's move on to the numerical analyses.
Here, we consider the bounds of the mass of $h$ and its $g_{ZZh}$ coupling.
The small $g_{ZZh}$ coupling means the large $g_{ZhA}$ coupling, so $hA$ production at LEP II can be enhanced.
However, if $m_h$ is larger than 90 GeV, then there is almost no constraint because of P-wave suppression \cite {PDG05}.
Therefore, we take $m_h>90$ GeV.
According to the Fig.~10 of \cite {Barate:2003sz}, 
the upper bounds on 
the $g_{ZZh}$ coupling, normalized by the SM coupling $g_{ZZ\varphi _{\rm SM}}$, should be
 about less than 0.5 ($\xi^2 \equiv g_{ZZh}^2/g_{ZZ\varphi _{\rm SM}}^2 \lesssim 0.25$) for $m_h>90$ GeV
 in 95\% confidence level\cite{Barate:2003sz}.
Therefore, 
the value of $\xi^2$ must be smaller than 0.25 through the following analyses.
When we identify the $2.3 \, \sigma $ excess at corresponding mass near $98$ GeV as a production signal of $h$, 
we take narrower region 95 GeV$<m_h<$101 GeV. The number of events observed is a tenth of the estimated number of the corresponding SM Higgs boson, which corresponds to the value of $\xi$ as $\xi^2 =0.1$.

Based on these setups, we have drawn three figures as Fig.\ref{inverse_a}-\ref{inverse_c}, which correspond
to three values of $A_t$ as $A_t$=300, 325, and 350 GeV, respectively.
Here we have used the on-shell top mass $m_t=175$ GeV 
and the one-loop potential 
in which $D$-term contribution to the sfermion masses is neglected\cite{Drees:1991mx}.
The horizontal axis is the CP-odd Higgs boson mass ($m_{A}$) and the vertical axis is $\tan\beta $.
%
Here, the doted-dashed lines represent contour lines of $m_h=90$, 95, and 101 GeV, respectively.
The dashed lines represent contour lines of $\xi^2$=0.05, 0.10, 0.15, and 0.25, respectively.
Thick solid lines represent contour lines of $m_H=$114.4, 115, and 120 and thin solid lines represent contour lines of $m_{H^\pm}=120$, 130, and 140 GeV, respectively. 
Each white area is the allowed region determined by the conditions of
$m_h>90$ GeV, $\xi^2<0.25$, and $m_H>114.4$ GeV.
Since top quark decays into charged Higgs boson and bottom quark in this case,
there are experimental upper bound for $\tan \beta$ ($\tan \beta \lesssim 50$  \cite{Abazov:2001md})\footnote{
Strictly speaking, the chargino mass bound $m_{\chi^\pm}>94$ GeV leads to the upper bound for $\tan\beta$
as $\tan\beta<23$ in the parameter set we took here. However, such bound can be easily weakened by taking
other parameter sets, so we do not take the bound seriously here.}.
When $A_t=325$ GeV, the allowed region shows that $m_h \lesssim 105$ GeV, $m_H \lesssim 118$ GeV, $92$ GeV $\lesssim m_A \lesssim 108$ GeV, $122$ GeV $\lesssim m_{H^{\pm }} \lesssim 134$ GeV, and $11 \lesssim \tan\beta \lesssim 50$. 
If we assume $2.3 \, \sigma $ excess as a signal of $h$, then we take
95 GeV$<m_h<$ 101 GeV as the allowed region.
This allowed region gives more constrained predictions that $m_H \lesssim 117$ GeV, $97$ GeV 
$\lesssim m_A \lesssim 103$ GeV, $126$ GeV $\lesssim m_{H^{\pm }} \lesssim 131$ GeV, and 
$20 \lesssim \tan\beta \lesssim 50$.


As a whole, 
we observe that there are parameter regions in which 
the lightest CP-even Higgs with sufficiently small $g_{ZZh}$ coupling and sufficiently large $m_H$ which
are consistent with LEP experiments are realized
with the natural parameter set.
Therefore this can be a solution for the little hierarchy problem.

Before ending this section, 
we explain the qualitative behavior of these figures.
At any given $m_A$, the smaller $\tan\beta$ leads to the larger $m_H$, the smaller
$m_h$, and the larger $g_{ZZh}$. 
This is because the off-diagonal entries of the neutral Higgs mass matrix, which becomes larger when
$\tan\beta$ is smaller, increase $m_H$, decrease $m_h$, and generate $g_{ZZh}$ coupling 
as discussed in the previous section.
On the contrary, for the fixed $\tan\beta$, the larger $m_A$ leads to the larger $m_H$, $m_h$,
$m_{H^\pm}$, and $g_{ZZh}$ coupling.  
This is because $m_A$ is the unique massive free parameter in the MSSM Higgs sector and 
the difference between the (1,1) and (2,2) components of the neutral Higgs mass matrix
becomes smaller when $m_A$ is larger, which leads to larger up-type Higgs component in the $h$
for the fixed off-diagonal component of the matrix. 

This scenario predicts a light charged Higgs whose mass is given as $m_{H^\pm}\sim 130$ GeV.
This is strongly disfavored from the constraint from the 
$b\rightarrow s \gamma $ process. This issue is a main subject of the next section.

\section{Constraint from $b\rightarrow s \gamma $ process}

We explore the consistency between the light charged Higgs which is required in this scenario and
the constraint from the $b\rightarrow s \gamma$ process.

As we have concluded in the previous section, 
the scenario with the small $g_{ZZh}$ coupling requires that all the MSSM Higgs bosons have the weak scale
masses. 
On the other hand, it is known that the lower limit of the charged Higgs mass  is strongly constrained 
by the $b\rightarrow s\gamma$ process since the experimental value of $ {\mathcal Br} (b\rightarrow s \gamma)_{exp.} = (355 \pm 24^{+9}_{-10} \pm 3) \times 10^{-6} $ \cite{unknown:2006bi} is now in good agreement with the SM prediction:
$ {\mathcal Br} (b\rightarrow s \gamma)_{\rm SM} = (360 \pm  30) \times 10^{-6} $ \cite{Gambino:2001ew,Buras:2002tp}.
For example, in the case of the type II 2HDM (the MSSM has almost the same Higgs sector), the lower bound 
is given by $m_{H^{\pm }} \geq 350$ GeV \cite{Gambino:2001ew} and this is greater than the mass derived in the previous section based on our scenario.
Such a severe constraint is caused by the fact that this additional charged Higgs induced amplitude 
always makes constructive contribution to the SM amplitude \cite{Ciuchini:1997xe}.
Therefore in this section we discuss this issue in detail.
Eventually we will find that the supersymmetric particles play an important role in solving this problem.



It is known that this lower bound for the charged Higgs mass can not be directly applied to that of the 
MSSM\cite{Ferrara:1974wb, Bertolini:1990if, Bertolini:1994cv, Barbieri:1993av}.
The reason is that there are additional contributions to this process induced by SUSY particles.
If the amplitudes induced by these SUSY particles contribute destructively enough to cancel the charged Higgs induced amplitudes in some parameter regions, then we can not limit the lower bound of charged Higgs mass.
Specifically, in the MSSM, the amplitude of $b\rightarrow s \gamma $ decay process is the sum of the following five different exchanges of intermediate particles.
\begin{enumerate}
\item W boson and up-type quarks

\item Charged Higgs boson and up-type quarks

\item Chargino and up-type squarks

\item Neutralino and down-type squarks

\item Gluino and down-type squarks

\end{enumerate}
The first two entries are the SM contribution and the charged Higgs contribution in the MSSM, 
respectively.
The third contribution is exactly the supersymmetric one of its first two contributions.
The remaining two contributions are caused by the off-diagonal elements of the down-type squark mass matrices in the basis where their fermionic partner's mass matrix is diagonalized.
These elements are induced by the renormalization group effect from the GUT to the weak scale even 
if we postulate flavor universal soft SUSY-breaking terms at the GUT scale.
In this case, however, it is known that the contributions of these neutralino and gluino induced amplitudes are not so significant compared to chargino contributions \cite{Bertolini:1990if}.
Therefore in the following analysis 
we ignore these last two categories.

We further make a few comments for the first three contributions before analyzing it numerically.
The essential point is that, in our scenario, the magnitudes of these amplitudes are almost the same order because all the particle in 
these loops have the weak scale masses. 
The charged Higgs mass must be around the weak scale to realize the small $g_{ZZh}$ coupling, 
and the chargino and
stop masses also must be around the weak scale because of our naturalness requirement. 
The next question is 
whether the chargino induced amplitude contributes to the other two amplitudes constructively or destructively.
To answer this issue, we consider the $b\rightarrow s\gamma$ process in exact SUSY limit.
In the $U(1)_{EM}$ SUSY gauge theory, 
the dimension five dipole type operator can not be written down in supersymmetric way \cite{Ferrara:1974wb}.
That is, if each supersymmetric particles have exactly the same mass spectrum as its own partners, 
then this radiative process must vanish.
This means that the contributions of the SM and charged Higgs induced amplitudes are completely compensated 
by its supersymmetric partners namely chargino contributions in exact SUSY limit.
Meanwhile, a number of authors have already pointed out that even in the case where SUSY is softly broken, there are parameter space which also make the chargino induced amplitude sufficiently destructive to the other two contributions \cite{Bertolini:1994cv}.
Thus, we expect that in our scenario, there will be some parameter regions which make the chargino induced amplitude destructive enough to compensate the charged Higgs induced amplitude to make whole our story consistent.
\begin{figure}
\begin{center}
\includegraphics[width=.6\textwidth]{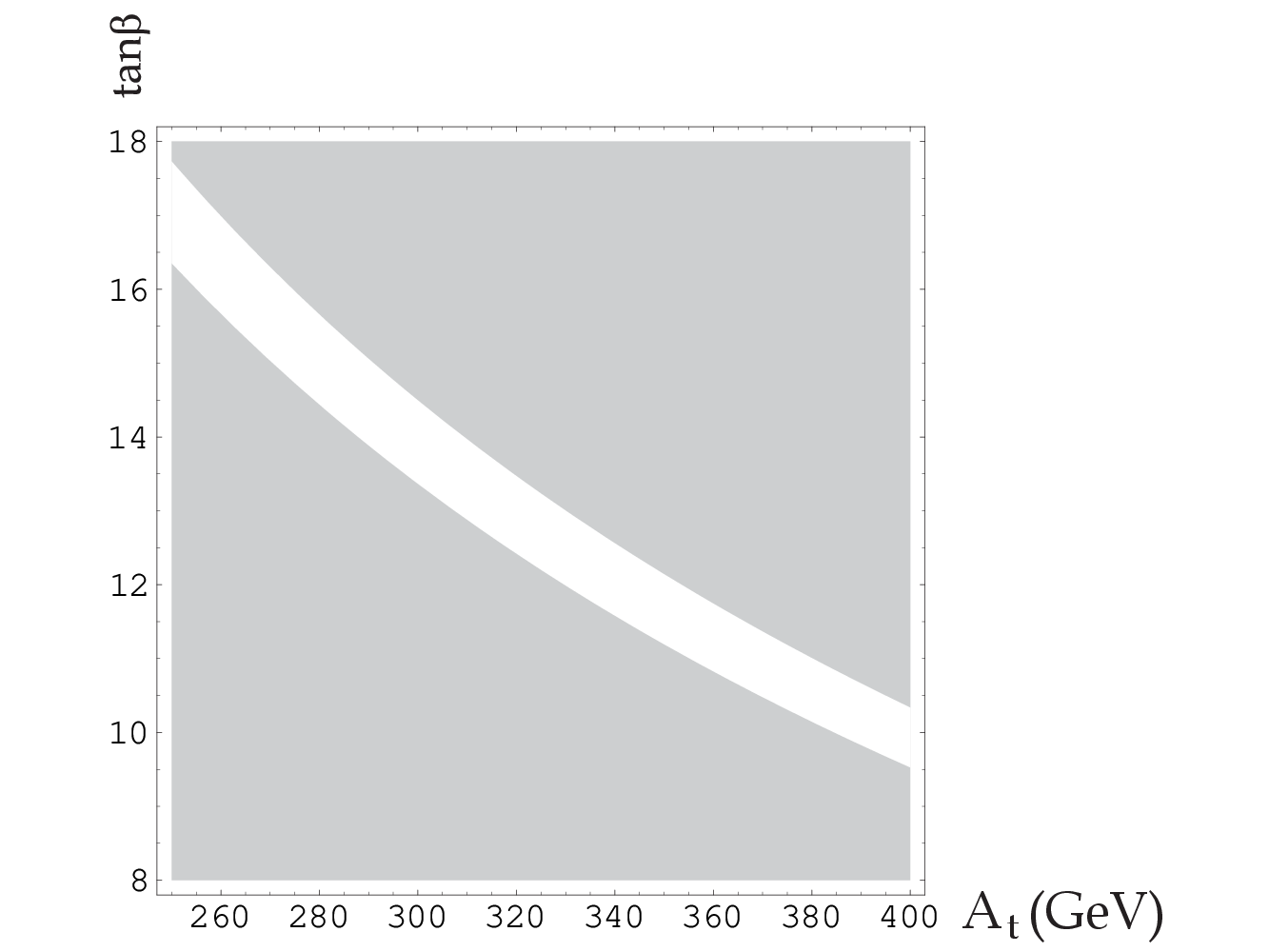}
\caption{ 
{\footnotesize
The white strip shows the parameter region for the suitable cancellation between the charged Higgs and chargino contributions to
$b\rightarrow s \gamma$ process. We required that the predicted value
of this model must be within one sigma deviation from the SM prediction. 
Here we fixed $m_{H^\pm}=125$ GeV 
and the other parameters were taken
as in the previous section. 
}
}
\label{a-tanbeta}
\end{center}
\end{figure}
\begin{figure}
\begin{center}
\includegraphics[width=.6\textwidth]{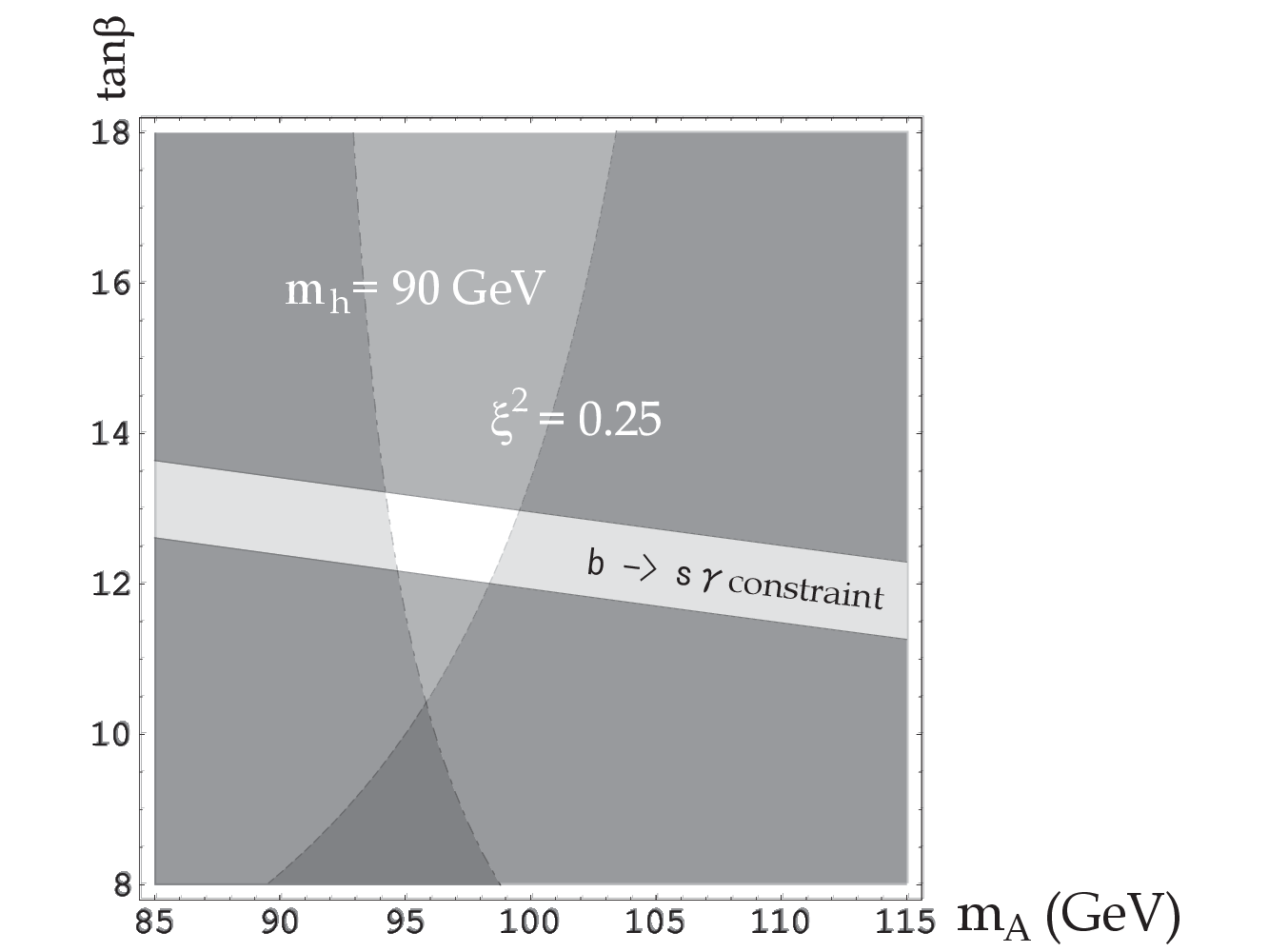}
\caption{ 
{\footnotesize
The cancellation region is plotted in $m_A$ and $\tan \beta$ plane at $A_t =325$ GeV.
This figure is superposed on the corresponding regions of Fig.\ref{inverse_b}.
Here we adopted the parameter set given in the previous section.
}}
\label{ma-tanbeta}
\end{center}
\end{figure}

Let's move on to the result of our numerical analyses shown in 
Fig.\ref{a-tanbeta} and Fig.\ref{ma-tanbeta}.
Fig.\ref{a-tanbeta} shows relations between $A_t$ and $\tan \beta$ required by the cancellations.
In this analysis we fixed charged Higgs boson mass as $m_{H^\pm }=125$ GeV and required that the sum of the 
charged Higgs and chargino induced amplitudes must be less than 5\% of that of the SM (This corresponds
to the requirement that the prediction of the scenario must be within one sigma deviation from the SM prediction). 
A white strip in Fig.\ref{a-tanbeta} shows the parameter regions 
which realize the suitable cancellations between the charged Higgs and the chargino contributions.
We see that 10\% tuning of $A_t$ is sufficient for the cancellation.
Fig.\ref{ma-tanbeta} shows relations between $m_A$ and $\tan \beta$ required by the cancellations at $A_t =325$ GeV.
This figure is superposed on the corresponding regions of Fig.\ref{inverse_b} and show that ${\mathcal Br} (b\rightarrow s \gamma )$ severely constrains allowed region of previous section. 

\section{Discussions and summary}
We examined a scenario in which the little hierarchy problem is solved by the lightest CP-even Higgs $h$ with
small $g_{ZZh}$ coupling. We showed that the LEP constraints can be satisfied even if the parameters which
contribute to the Higgs mass parameters are taken to be order of the weak scale. The essential point is simple.
The heavier CP-even Higgs $H$ becomes heavier by off-diagonal components of CP-even Higgs mass matrix, which
is important to satisfy the LEP bound for the SM Higgs because the heavier CP-even Higgs has almost the same
$g_{ZZH}$ coupling as the SM Higgs. On the other hand, the lighter CP-even Higgs $h$ becomes lighter by
the off-diagonal components, but the Higgs is not produced in LEP as much as the SM Higgs because it has only 
the small $g_{ZZh}$ coupling.

One of the interesting results in the scenario is that all the Higgs should have the weak scale masses. Actually, 
the charged Higgs mass must be around 130 GeV. However, this looks inconsistent with the constraint from the 
$b\rightarrow s\gamma$ process, $m_{H^\pm}>350$ GeV. We showed that the naturalness requirement for the 
parameters leads to the cancellation between the charged Higgs contribution and the chargino contribution
and the parameter space exists in which the scenario is consistent with the $b\rightarrow s\gamma$
constraint. Of course, if other contributions, neutralino or gluino contributions, become sizable, then
the parameter region may shift. So the predicted region in the $\tan\beta$ and $A_t$ plane 
may not have to be taken so seriously. 
What is important here is that such cancellation which is consistent with the SM prediction within one standard
deviation can be realized by only 10\% tuning of a parameter $A_t$ at the weak scale. This tuning becomes much
milder at the GUT scale to be almost 100\% tuning of the parameter $A$ at the GUT scale.

Other constraints, (e.g. from $K$ and $B$ meson mixings, precision electroweak parameters, etc.) than from
$b\rightarrow s\gamma$ process might give another severe constraints in principle. However, we expect that
such constraints are not so severe. One of the reason is that the couplings between
  the top quark and  the charged Higgs are suppressed because the most 
 of up-type Higgs is absorbed by the Higgs mechanism and the main component of the physical
  charged Higgs comes from the down-type Higgs if $\tan\beta>1$.  Moreover, for $K$, $B_d$ and $B_s$ meson mass mixings, the charged Higgs and the chargino contributions produce mainly the operator which includes only left-handed down type quarks as in the SM. The constraint from the operator is much weaker than from the operators which include both left and right handed down-type quarks.

Processes to which the light charged Higgs contribute in tree level may be more important, because
there are no contributions from SUSY partners. 
Since the charged Higgs couples more strongly 
with the third generation fields, so the processes which include the third generation fields are more preferable. 
The top decay process $t\rightarrow b+H^\pm$ is one of the candidates, but unfortunately the constraints
from Tevatron experiments are not severe. They rejected only very large $\tan\beta>50$ region or very 
small $\tan\beta<1$ region.  This is because main component of up-type Higgs are absorbed
by the Higgs mechanism and the main component of the physical charged Higgs comes from the down-type
Higgs if $\tan\beta>1$. In the $B$ decay, $B\rightarrow \tau\nu_\tau$ and 
$b\rightarrow c H^{\pm*}\rightarrow c \tau\nu_\tau$ are the candidate processes. Especially, the 
$B\rightarrow\tau\nu_\tau$ process is important because the SM contribution is suppressed by the chirality.
The branching ratio is given by
\begin{equation}
Br(B^-\rightarrow \tau^-\bar\nu_\tau)_{SM+CH}=Br(B^-\rightarrow\tau^-\bar\nu_\tau)_{SM}\times r_H,\quad
r_H=\left(1-m_B^2\frac{\tan^2\beta}{m_{H^\pm}^2}\right)^2,
\end{equation}
where $m_B$ is the $B$ meson mass and $Br(B^-\rightarrow\tau^-\bar\nu_\tau)_{SM}=(1.59\pm 0.40)\times 10^{-4}$ \cite{Ikado:2006un}
 is the SM prediction
for the branching ratio \cite{Hou:1992sy}. 
Recently, the first evidence for the process was reported by Belle 
$( Br(B^+\rightarrow\tau^+\nu_\tau)=(1.79^{+0.56+0.39}_{-0.49-0.46})
\times 10^{-4})$ \cite{Browder:ICHEP2006} and 
Babar has also reported as 
$( Br(B^+\rightarrow\tau^+\nu_\tau)=
(0.88^{+0.68+0.11}_{-0.67-0.11})\times 10^{-4})$\cite{babar}. The combined branching ratio 
$( Br(B^+\rightarrow\tau^+\nu_\tau)=(1.34\pm 0.48)\times 10^{-4})$ gives us 
$r_H=0.86\pm 0.37$ and 
this is consistent with the SM prediction
. This measurement has already restricted 
the  parameter 
space as $\tan\beta<20$ and $29<\tan\beta<37$ (95\% C.L.) when $m_{H^\pm}\sim 130$ GeV. 
However, in our scenario, the smaller $\tan\beta$ seems to be favored so that plenty of parameter space is still alive.
We expect more precise experimental value and theoretical calculation to see the deviation from the SM in 
this decay.
The $b\rightarrow c H^{\pm*}\rightarrow c \tau\nu_\tau$ process may be useful\cite{Tanaka:1994ay, Akeroyd:2004mj} if it is detected, 
but it may be difficult to reject the intermediate $\tan\beta\sim 10$ region. 


In this scenario, the lightest SUSY particle is the Bino unless the very weakly coupled SUSY particle like
the gravitino or axino are lighter than the Bino. Since in our scenario, we do not have to take care
about the constraint from the $b\rightarrow s\gamma$ process, so the bulk region in the parameter space for the
dark matter may survive. So the analysis for the dark matter must be re-considered, but this issue is
beyond the scope of this paper. 

In this paper, we have used several assumptions to calculate the Higgs masses and $b\rightarrow s \gamma$ amplitude.
For example, we adopted the GUT relation for the gaugino masses. And of course, we could calculate the Higgs masses 
with more precise approximation. However, we would like to emphasize that our arguments for weakening LEP bound for
the Higgs mass and canceling $b\rightarrow s\gamma$ amplitude are quite general one and the scenario in which 
the lighter CP-even 
Higgs have small $g_{ZZh}$ coupling is an interesting possibility to solve the little hierarchy problem. We hope 
that this scenario will be tested in future experiments, LHC or ILC or (super-)B factory etc \cite{Cao:2003tr}.


\section*{Acknowledgement}
N.M. thanks K. Tobe and the organizers of SI2005 in which he obtained an opportunity to discuss with K. Tobe 
on the invisible Higgs.
N.M., K.S., and A.I.S. are supported in part by Grants-in-Aid for Scientific 
Research from the Ministry of Education, Culture, Sports, Science, 
and Technology of Japan.
The work of T.Y. was
supported by 21st Century COE Program of Nagoya University provided by JSPS.




\begin{thebibliography}{99}

%
%
\bibitem{Barate:2003sz}
  R.~Barate {\it et al.}  [LEP Working Group for Higgs boson searches],
  Phys.\ Lett.\ B {\bf 565}, 61 (2003)
  [arXiv:hep-ex/0306033].

%
%

\bibitem{MSSMHiggs}
  Y.~Okada, M.~Yamaguchi, and M.~Yanagida,
  Prog. Theor. Phys. Lett. {\bf 85}, 1(1990);
  J.~Ellis, G.~Ridolfi, and F.~Zwirner,
  Phys. Lett. B {\bf 257} 83, (1991); {\bf 262} 477 (1991).


\bibitem{little1}
 M.~Bastero-Gil, C.~Hugonie, S.F.~King, D.P.~Roy, and S.~Vempati,
 Phys. Lett. B {\bf 489}, 359 (2000) [arXiv:hep-ph/0006198];
 K.~Agashe and M.~Graesser, Nucl. Phys. B {\bf 507}, 3 (1997)
 [arXiv:hep-ph/9704206].
 
\bibitem{little2}
 J.L.~Feng, K.T.~Matchev, and T.~Moroi, Phys. Rev. Lett. {\bf 84},
 2322 (2000) [arXiv:hep-ph/9908309]; Phys. Rev. D {\bf 61}, 075005
 (2000) [arXiv:hep-ph/9909334].
 
\bibitem{little3}
 A.~Brignole, J.A.~Casas, J.R.~Espinosa, and I.~Navarro, Nucl. Phys.
 B {\bf 666}, 105 (2003) [arXiv:hep-ph/0301121]; J.A.~Casas, J.R.~Espinosa,
 and I.~Hidalgo, JHEP {\bf 0401}, 008 (2004) [arXiv:hep-ph/0310137].
 
\bibitem{little4}
 P.~Batra, A.~Delgado, D.E.~Kaplan, and T.M.P.~Tait, JHEP {\bf 0402}, 043 (2004)
 [arXiv:hep-ph/0309149]; JHEP {\bf 0406}, 032 (2004) [arXiv:hep-ph/0404251];
 A.~Maloney, A. Pierce, and J.G.~Wacker, arXiv:hep-ph/0409127.
 
\bibitem{little5}
 R.~Harnik, G.D.~Kribs, D.T.~Larson, and H.~Murayama, Phys. Rev. D {\bf 70},
 015002 (2004) [arXiv:hep-ph/0311349]; S.~Chang, C.~Kilic, and R.~Mahbubani,
 Phys. Rev. D {\bf 71}, 015003 (2005) [arXiv:hep-ph/0405267]; A.~Birkedal, Z.~Chacko,
 and Y.~Nomura, Phys. Rev. D {\bf 71}, 015006 (2005) [arXiv:hep-ph/0408329];
 A.~Delgado and T.M.P.~Tait, JHEP {\bf 0507}, 023 (2005) [arXiv:hep-ph/0504224].
 
\bibitem{little6}
 T.~Kobayashi and H.~Terao, JHEP {\bf 0407}, 026 (2004) [arXiv:hep-ph/0403298];
 T.~Kobayashi, H.~Nakano, and H.~Terao, Phys. Rev. D {\bf 71}, 115009 (2005)
 [arXiv:hep-ph/0502006].

\bibitem{little7}
 A.~Birkedal, Z.~Chacko, and M.K.~Gaillard, JHEP {\bf 0410}, 036 (2004) [arXiv:hep-ph/0404197]:
 P.H.~Chankowski, A.~Falkowski, S.~Pokorski, and J.~Wagner, Phys. Lett. B {\bf 598}, 252
 (2004) [arXiv:hep-ph/0407242]; Z.~Berezhiani, P.H.~Chankowski, A.~Falkowski, and S.~Pokorski, Phys. Rev. Lett. {\bf 96}, 031801 (2006) 
 [arXiv:hep-ph/0509311]; T.~Roy and M.~Schmaltz, JHEP {\bf 0601}, 149 (2006) [arXiv:hep-ph/0509357]; C.~Csaki, G.~Marandella,
 Y.~Shirman, and A.~Strumia, Phys. Rev. D {\bf 73}, 035006 (2006)  [arXiv:hep-ph/0510294].

\bibitem{little8}
 R.~Dermisek and J.F.~Gunion, Phys. Rev. Lett. {\bf 95}, 041801 (2005) [arXiv:hep-ph/0502105];
 S.~Chang, P.J.~Fox, and N.~Weiner, JHEP {\bf 08}, 068 (2006) [arXiv:hep-ph/0511250]; P.C.~Schuster and N.~Toro,
 [arXiv:hep-ph/0512189].
 
\bibitem{little9}
 Z.~Chacko, Y.~Nomura, and D.~Tucker-Smith, Nucl. Phys. B {\bf 725}, 207 (2005)
 [arXiv:hep-ph/0504095]; Y.~Nomura and B.~Tweedie, Phys. Rev. D {\bf 72}, 015006
 (2005) [arXiv:hep-ph/0504246].

\bibitem{little10}
 K. Choi, K.S.~Jeong, T.~Kobayashi, and K.I.~Okumura, Phys. Lett. B {\bf 633}, 355 (2006)
 [arXiv:hep-ph/0508029].
  
\bibitem{little11}
 R.~Kitano and Y.~Nomura, Phys. Lett. B {\bf 631}, 58 (2005) [arXiv:hep-ph/0509039].

 
\bibitem{little12}
 Y.~Nomura, D.~Poland, and B.~Tweedie, Nucl. Phys. B {\bf 745}, 29 (2006) [arXiv:hep-ph/0509243];Phys. Lett. B {\bf 633}, 573 (2006) [arXiv:hep-ph/0509244].
 
 
\bibitem{little13}
 R.~Dermisek and H.D.~Kim, Phys. Rev. Lett. {\bf 96}, 211803 (2006) [arXiv:hep-ph/0601036].
 
 
\bibitem{Kane:2004tk}
  G.~L.~Kane, T.~T.~Wang, B.~D.~Nelson, and L.~T.~Wang,
  Phys.\ Rev.\ D {\bf 71}, 035006 (2005)
  [arXiv:hep-ph/0407001].

\bibitem{Drees:2005jg}
  M.~Drees,
  Phys.\ Rev.\ D {\bf 71}, 115006 (2005)
  [arXiv:hep-ph/0502075].


\bibitem{Gambino:2001ew}
  P.~Gambino and M.~Misiak,
  Nucl.\ Phys.\ B {\bf 611}, 338 (2001)
  [arXiv:hep-ph/0104034].

%
%

\bibitem{PDG05} S. Eidelman et al., [Particle Data Group], Phys. Lett. B 592, 1 (2004) and 
2005 partial update for the 2006 edition available on the PDG WWW pages (URL: http://pdg.lbl.gov/); 
W.-M. Yao et al., [Particle Data Group], J. Phys. G{\bf 33}, 1
(2006). 




\bibitem{Drees:1991mx}
  M.~Drees and M.~M.~Nojiri,
  Phys.\ Rev.\ D {\bf 45}, 2482 (1992).




%

\bibitem{Abazov:2001md}
  V.~M.~Abazov {\it et al.}  [D0 Collaboration],
  Phys.\ Rev.\ Lett.\  {\bf 88}, 151803 (2002)
  [arXiv:hep-ex/0102039];
  A.~Abulencia {\it et al.}  [CDF Collaboration],
  Phys.\ Rev.\ Lett.\  {\bf 96}, 042003 (2006)
  [arXiv:hep-ex/0510065].



%
%

\bibitem{unknown:2006bi}
    Heavy Flavor Averaging Group (HFAG),
  [arXiv:hep-ex/0603003],  http://www.slac.stanford.edu/xorg/hfag/ ~~. 


%
%


%
%

\bibitem{Buras:2002tp}
  A.~J.~Buras, A.~Czarnecki, M.~Misiak, and J.~Urban,
  Nucl.\ Phys.\ B {\bf 631}, 219 (2002)
  [arXiv:hep-ph/0203135];
  A.~J.~Buras and M.~Misiak,
  Acta Phys.\ Polon.\ B {\bf 33}, 2597 (2002)
  [arXiv:hep-ph/0207131];
  T.~Hurth,
  Rev.\ Mod.\ Phys.\  {\bf 75}, 1159 (2003)
  [arXiv:hep-ph/0212304].


%
%


\bibitem{Ciuchini:1997xe}
  M.~Ciuchini, G.~Degrassi, P.~Gambino, and G.~F.~Giudice,
  Nucl.\ Phys.\ B {\bf 527}, 21 (1998)
  [arXiv:hep-ph/9710335];
F.~M.~Borzumati and C.~Greub,
  Phys.\ Rev.\ D {\bf 58}, 074004 (1998)
  [arXiv:hep-ph/9802391];
  F.~M.~Borzumati and C.~Greub,
  Phys.\ Rev.\ D {\bf 59}, 057501 (1999)
  [arXiv:hep-ph/9809438];
  F.~Borzumati, C.~Greub, and Y.~Yamada,
  Phys.\ Rev.\ D {\bf 69}, 055005 (2004)
  [arXiv:hep-ph/0311151].

%
%
\bibitem{Ferrara:1974wb}
  S.~Ferrara and E.~Remiddi,
  Phys.\ Lett.\ B {\bf 53}, 347 (1974).




%
%


\bibitem{Bertolini:1990if}
  S.~Bertolini, F.~Borzumati, A.~Masiero, and G.~Ridolfi,
  Nucl.\ Phys.\ B {\bf 353}, 591 (1991).

%
%

\bibitem{Bertolini:1994cv}
  S.~Bertolini and F.~Vissani,
  Z.\ Phys.\ C {\bf 67}, 513 (1995)
  [arXiv:hep-ph/9403397].

%
%

\bibitem{Barbieri:1993av}
  R.~Barbieri and G.~F.~Giudice,
  Phys.\ Lett.\ B {\bf 309}, 86 (1993)
  [arXiv:hep-ph/9303270];
  N.~Oshimo,
  Nucl.\ Phys.\ B {\bf 404}, 20 (1993);
  M.~A.~Diaz,
  Phys.\ Lett.\ B {\bf 304}, 278 (1993)
  [arXiv:hep-ph/9303280];
  Y.~Okada,
  Phys.\ Lett.\ B {\bf 315}, 119 (1993)
  [arXiv:hep-ph/9307249];
  R.~Garisto and J.~N.~Ng,
  Phys.\ Lett.\ B {\bf 315}, 372 (1993)
  [arXiv:hep-ph/9307301];
  V.~D.~Barger, M.~S.~Berger, P.~Ohmann, and R.~J.~N.~Phillips,
  Phys.\ Rev.\ D {\bf 51}, 2438 (1995)
  [arXiv:hep-ph/9407273];
  T.~Goto and Y.~Okada,
  Prog.\ Theor.\ Phys.\  {\bf 94}, 407 (1995)
  [arXiv:hep-ph/9412225];
  M.~Ciuchini, G.~Degrassi, P.~Gambino, and G.~F.~Giudice,
  Nucl.\ Phys.\ B {\bf 534}, 3 (1998)
  [arXiv:hep-ph/9806308];
  F.~Borzumati, C.~Greub, T.~Hurth, and D.~Wyler,
  Phys.\ Rev.\ D {\bf 62}, 075005 (2000)
  [arXiv:hep-ph/9911245];
  G.~Degrassi, P.~Gambino, and G.~F.~Giudice,
  JHEP {\bf 0012}, 009 (2000)
  [arXiv:hep-ph/0009337];
  T.~Besmer, C.~Greub, and T.~Hurth,
  Nucl.\ Phys.\ B {\bf 609}, 359 (2001)
  [arXiv:hep-ph/0105292];
  A.~J.~Buras, P.~H.~Chankowski, J.~Rosiek, and L.~Slawianowska,
  Nucl.\ Phys.\ B {\bf 659}, 3 (2003)
  [arXiv:hep-ph/0210145];
  M.~Ciuchini, E.~Franco, A.~Masiero, and L.~Silvestrini,
  Phys.\ Rev.\ D {\bf 67}, 075016 (2003)
  [Erratum-ibid.\ D {\bf 68}, 079901 (2003)]
  [arXiv:hep-ph/0212397];
  K.~i.~Okumura and L.~Roszkowski,
  JHEP {\bf 0310}, 024 (2003)
  [arXiv:hep-ph/0308102];
  J.~Foster, K.~i.~Okumura, and L.~Roszkowski,
  JHEP {\bf 0603}, 044 (2006)
  [arXiv:hep-ph/0510422];
  G.~Degrassi, P.~Gambino, and P.~Slavich,
  Phys.\ Lett.\ B {\bf 635}, 335 (2006)
  [arXiv:hep-ph/0601135].







%
%

\bibitem{Hou:1992sy}
  W.~S.~Hou,
  Phys.\ Rev.\ D {\bf 48}, 2342 (1993).


%
%


\bibitem{Browder:ICHEP2006}
T. Browder, talk presented at the 33rd International Conference on 
High Energy Physics (ICHEP2006), 2006, Moscow.   


\bibitem{Ikado:2006un}
  We use same input parameters quoted in the following article.
  K.~Ikado {\it et al.},
  arXiv:hep-ex/0604018.

%
%

\bibitem{babar}
Babar collaboration, [arXiv:hep-ex/0608019]; see also talk presented by S.J.~Sekula
at ICHEP06, Moscow.

%
%

\bibitem{Tanaka:1994ay}
  M.~Tanaka,
  Z.\ Phys.\ C {\bf 67}, 321 (1995)
  [arXiv:hep-ph/9411405];
  T.~Miki, T.~Miura, and M.~Tanaka,
  arXiv:hep-ph/0210051.

%
%

\bibitem{Akeroyd:2004mj}
  A.~G.~Akeroyd {\it et al.}  [SuperKEKB Physics Working Group],
  arXiv:hep-ex/0406071.






%
%

\bibitem{Cao:2003tr}
  See, e.g.,
  Q.~H.~Cao, S.~Kanemura and C.~P.~Yuan,
  Phys.\ Rev.\ D {\bf 69}, 075008 (2004)
  [arXiv:hep-ph/0311083],
  A.~Belyaev, Q.~H.~Cao, D.~Nomura, K.~Tobe and C.~P.~Yuan,
  arXiv:hep-ph/0609079.


\end{thebibliography}
\end{document}